# A Blind-Mixing Scheme for Bitcoin based on an Elliptic Curve Cryptography Blind Digital Signature Algorithm

QingChun ShenTu[1*], JianPing Yu [1]

[1] ATR Defense Science & Technology Lab., Shenzhen University, Shenzhen, China
* unshadowster@gmail.com

**Abstract:** To strengthen the anonymity of Bitcoin, several centralized coin-mixing providers (mixers) such as BitcoinFog.com, BitLaundry.com, and Blockchain.info assist users to mix Bitcoins through CoinJoin transactions with multiple inputs and multiple outputs to uncover the relationship between them. However, these mixers know the output address of each user, such that they cannot provide true anonymity. This paper proposes a centralized coin-mixing algorithm based on an elliptic curve blind signature scheme (denoted as Blind-Mixing) that obstructs mixers from linking an input address with an output address. Comparisons among three blind signature based algorithms, Blind-Mixing, BlindCoin, and RSA Coin-Mixing, are conducted. It is determined that BlindCoin may be deanonymized because of its use of a public log. In RSA Coin-Mixing, a user's Bitcoins may be falsely claimed by another. In addition, the blind signature scheme of Blind-Mixing executes 10.5 times faster than that of RSA Coin-Mixing.

**Keywords:** Bitcoin, anonymity, Blind Signature, Coin-Mixing, Blind-Mixing

## 1. Introduction

Bitcoin is a decentralized crypto-currency that was introduced by Nakamoto [1] in 2008, and was deployed in January 2009. Bitcoin has several characteristics including peer-to-peer protocols, decentralized production of Bitcoins by the proof of work (PoW) protocol, the prevention of double spending by transparent transactions, pseudo-anonymity, and personal privacy protection, which have made Bitcoin increasingly popular. Its market value reached at peak of $12 billion in 2013.

In the Bitcoin network, Bitcoin addresses act as user accounts. Generally speaking, the aim of anonymization is to prevent attackers from discovering the relationship between Bitcoin addresses and real or virtual user identity information through the Bitcoin network and the blockchain used to record transactions. Conversely, deanonymization is the uncovering of the relationship between the Bitcoin address and the user in which the IP address is important user identity information.

Research regarding Bitcoin coin-mixing originated from the CoinJoin anonymization method proposed by Gmaxwell [2]. Current study regarding coin-mixing focuses on three directions. The first is the use of centralized websites, such as BitcoinFog.com, BitLaundry.com, and Blockchain.info, where Mixcoin [3] was a representative study. The second regards decentralized protocols such as Fair Exchange [4], XIM [5], CoinParty [6], and CoinShuffle [7], which are compatible with the Bitcoin protocol, where Bitcoins are mixed through decentralized protocols, and no trust is required among users with no possibility of currency loss. The last direction regards novel coin-mixing technologies such as Zerocoin [8], Zerocash [9], Pinocchio [10], and CryptoNote [11], which are not compatible with the Bitcoin Protocol, and must be applied in new blockchains.

This paper discusses centralized coin-mixing for Bitcoin, where a serious weakness of centralized coin-mixing is that coin-mixing providers (mixers) must know a user's input address and output address so that Bitcoins can be mixed correctly and currency can be correctly output to a user's output address. Hence, centralized coin-mixing cannot provide true anonymity for users.

A blind digital signature represents a condition where a signer signs the digest of a message while the content of the message is unknown to the signer. In 1982, David Chaum [12] first proposed to implement an anonymous e-Cash system based on the use of blind signatures, which was intended to protect the anonymity of a sender unconditionally. In 1993, Okamoto [13] proposed the first blind signature scheme based on the discrete logarithm. In 2012, Watson Ladd [14] proposed making Bitcoin transactions untraceable using blind signature. In 2015, BlindCoin [15] was proposed based on bilinear groups to make centralized coin-mixing more anonymous. In July 2015, Wu [16] also proposed a blind coin-mixing algorithm based on RSA (RSA Coin-Mixing). The present study intends to avoid several weaknesses associated with BlindCoin and RSA Coin-Mixing, and proposes a practical coin-mixing algorithm that supports more complete anonymization.

## 2. Background

### 2.1. Transactions and the transaction chain

A Bitcoin address is an account on the Bitcoin network, which corresponds to a bank account in conventional currency systems. A Bitcoin address is generated by double hashing a public key. Only the user who owns the corresponding private key can make use of Bitcoins lodged at this address.

The Bitcoin system has a public ledger that stores transfer records rather than the balance of every Bitcoin address. A transfer record is known as a Bitcoin transaction, and it includes the transfer time, inputs, outputs, amounts, and signatures. Bitcoin transactions with their related inputs and outputs enter into the transaction chain, as illustrated in Fig. 1. A page in the public ledger is denoted as a block. Transactions being generated during the latest ten minutes are recorded in the latest block, and we regard these transactions as being acknowledged once. N blocks are generated subsequent to the first block, and then these transactions will be acknowledged N + 1 times.

Every transaction has a unique identification (ID). Each input is connected with the output of the previous transaction so that the input address of a transaction can be obtained through the output address of the previous transaction. All inputs should not be used within all existing transactions because this will prevent the successful verification of the transaction. Signatures aim to prove that an input amount belongs to the sender because only the private key owner can sign the transaction properly.

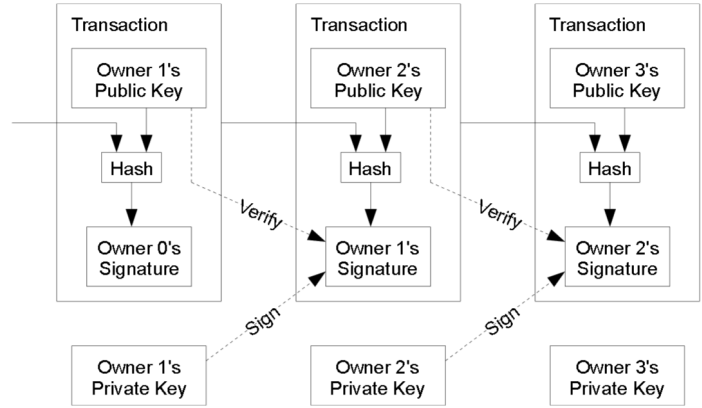

***Fig. 1*** *Bitcoin transactions and the chain of transactions*

*2.2. Elliptic curve cryptography system used by Bitcoin*

Bitcoin adopts the ECC system as its signature algorithm, and its elliptic curve is secp256k1 [17], whose formation is

$$y^2 \equiv x^3 + ax + b \pmod{p} \quad . \tag{F1}$$

This curve can be described as T = (p, a, b, G, n, h), where a and b are constants, p is the p value of the finite field F(p) of secp256k1, G is the base point, n is the order of G, and h is a cofactor. Their values are listed as follows:

a = 0, b = 7, h = 1, p = $2^{256} - 2^{32} - 2^9 - 2^8 - 2^7 - 2^6 - 2^4 - 2^1$,
G = 02 79BE667E F9DCBBAC 55A06295 CE870B07 029BFCDB 2DCE28D9 59F2815B 16F81798,
n = FFFFFFFF FFFFFFFF FFFFFFFF FFFFFFFE BAAEDCE6 AF48A03B BFD25E8C D0364141.

The public key and private key generating process can be outlined as follows.

(1) Select a large integer d < p, and d ∈ F(p);

(2) Calculate P = dG;

(3) Obtain the key pair (P, d), where P is the public key and d is the private key**.**

*2.3. The blind signature scheme of ECC*

In the general signature scheme illustrated in Fig. 2(a), the signer produces a digital signature on known message content. Compared to the general signature scheme, the process of blind signature [12] is as illustrated in Fig. 2(b). The requester performs a blinding shift on the message, and sends the blinded message to the signer. The signer signs the blinded message and returns the blind signature to the requester, and then the requester unblinds the blind signature to obtain the signature on the original message.

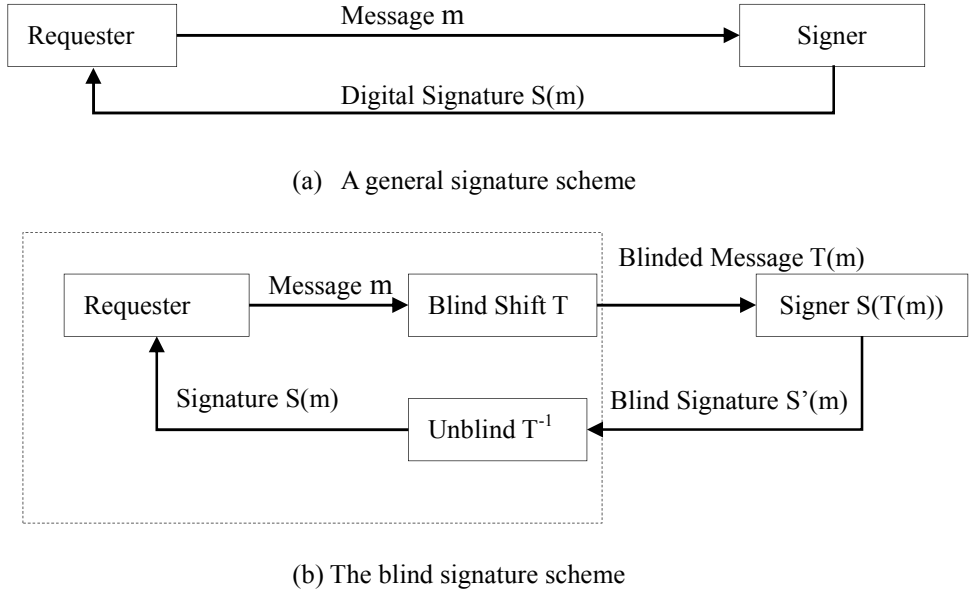

(a) A general signature scheme

(b) The blind signature scheme

*Fig. 2* Comparison between a general signature scheme and the blind signature scheme

The blind signature scheme of ECC adopts the same secp256k1 elliptic curve as Bitcoin. Suppose that the key pair of the signer is (P, d), the message is m, and all else is as has been defined. The blind signature algorithm from Zhang [19] is as follows.

(1) The signer randomly selects an integer $k \in Z_n^*$, calculates R = kG, and then transmits R to the requester.
(2) The requester randomly selects two integers $\gamma$ and $\delta \in Z_n^*$, blinds the message, and then calculates point A = kG + $\gamma$G + $\delta$P = (x, y), t = x (mod n). If t equals zero, then $\gamma$ and $\delta$ should be reselected. The requester calculates c = SHA256 (m || t), c' = c − $\delta$, where SHA256 is a novel hash function computed with 32-bit words and c' is the blinded message, and then sends c' to the signer.
(3) The signer calculates the blind signature s' = k − c'd, and then sends it to the requester.
(4) The requester calculates s = s' + $\gamma$, and (c, s) is the signature on m.
(5) Both the requester and signer can verify the signature (c, s) through the formation
   c = SHA256(m || $R_x$(cP + sG) mod n),                     (F2)

where $R_x(cP + sG)$ represents obtaining the x resolution values of point $cP + sG$, and·||·represents the concatenation of two strings.

A blind signature algorithm has also been developed based on RSA. ECC demands a key size of 160 bits for security that is equivalent to that obtained by a 1024-bit RSA key size [19]. A shorter key size requires less storage space and reduced computation. Therefore, we choose ECC as the cryptography algorithm of Blind-Mixing. In addition, for user convenience, Blind-Mixing also makes use of the Part Blind concept proposed by Abe [20], where some information confirmed in advance, such as amount, validity period, and output address, are inserted into the message waiting to be signed.

## 3. Blind-Mixing

### 3.1. Application scheme

We construct an application scheme as follows. Suppose that there exists a Bitcoin bank, and a user deposits Bitcoins in the bank, withdraws Bitcoins from the bank, and transfers Bitcoins to businesses via the bank. Businesses can obtain payment from the bank after the bank verifies the withdrawal voucher. The scheme is illustrated in Fig. 3.

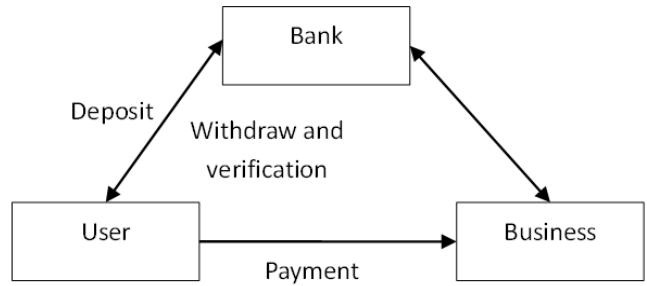

*Fig. 3* The application scheme of Blind-Mixing

The detailed operations that can be performed by the user are given as follows.

(1) Deposit: the user sends Bitcoins to the deposit address obtained from the bank, and generates a deposit voucher to certify the deposit. The bank verifies the deposit voucher and performs blind signature. The user receives and unblinds the blind signature and obtains the true signature. We denote the signature plus the original message as the withdrawal voucher.

(2) Payment: the user can transfer the withdrawal voucher to a business, and the business obtains the verified result from the bank and accepts the withdrawal voucher.

(3) Verification: the user requests verification of the withdrawal voucher and the bank returns the verification result.

(4) Withdrawal: the user requests a withdrawal from the bank through the withdrawal voucher, and the bank sends Bitcoins to the output address of the user after verification.

*3.2. Attacker Models*

Several attacker models affecting the anonymity of Blind-Mixing can be described as follows.

***Passive attacker model***: the attacker passively analyses the output of Bitcoins through the transaction chain of the Bitcoin system.

***Active attacker model***: in addition to capabilities of the passive attacker, the active attacker also obeys the Blind-Mixing protocol and engages in the coin-mixing process as a normal user. Furthermore, the attacker is able to cheat the Blind-Mixing protocol but not damage the protocol.

***Super attacker model***: in addition to capabilities of the active attacker, the super attacker can sniff all communication data between the user and the bank, and also controls the bank. However, the attacker aims at deanonymizing users rather than interrupting the normal process of coin-mixing.

*3.3. Blind-Mixing*

Blind-Mixing runs on the TOR [21] anonymity network to encrypt communication data between the user and the bank, and to hide the real IP addresses employed by the users and banks, avoiding sniffer attacks. The protocol can also simultaneously prevent banks from linking input and output addresses through a user's real IP address.

**Step 1. System initialization**

(1) The bank generates a key pair (P, d), makes the public key public, and retains the private key.

(2) Define a series of allowable denominations $v_i = \{v_1, v_2, \ldots v_n\}$, i = 1, 2, … n, that users must employ when performing coin-mixing.

(3) Define the message to be signed m = {O || $v_i$ || P || nonce}, where $O$ represents the output address, $v_i$ is the withdrawal amount, $P$ refers to the public key of the bank, which is used to identify different banks, and *nonce* is a random integer used for generating different messages.

(4) Define the required number of acknowledgements, where, after the deposit transaction (DT) obtains the specific number of acknowledgements in the Bitcoin system, the bank will perform the blind signature.

(5) Define the withdrawal delay, which represents the length of time required for the user to withdraw Bitcoins after completing a DT, which is normally 1 to 3 days, to ensure that Bitcoins are adequately mixed.

(6) The withdrawal address represents the source address of the withdrawal transaction. The bank generates a series of withdrawal addresses, and sends a certain amount Bitcoins to them when preparing for the withdrawal requested by the user.

**Step 2. User obtains the public key of the bank and makes deposits**

Suppose the amount of coin-mixing by the bank requested by the user is $v_i$. The bank generates a new key pair (R, k), sends the public key of the deposit address and the signature of R (R, $sign_d(R)$) to the user, and the bank makes use of this address to receive Bitcoins from the user. The user obtains the deposit address based on R. The user verifies the public key and the signature. If the signature is not authentic, then an error is returned.

The user sends Bitcoins to the deposit address, which is denoted as a deposit, and records the transaction ID tx_id, the entire transaction of which is denoted as a deposit transaction (DT). The user's input address is assigned as the input address of the DT, where the corresponding key pair is (Q, f). The user must ensure that a sufficient amount of Bitcoins has been lodged at the input address in advance, which is verified by the bank.

**Step 3. User blinds the message**

For m = {O || $v_i$ || P || nonce}, the user selects two integers γ and δ ∈ $Z_n^*$, and calculates:

$$A = kG + \gamma G + \delta P = (x, y), \quad t = x \pmod{n}, \quad c = SHA256(m \parallel t), \quad c' = c - \delta. \tag{F3}$$

The above process is denoted as blinding. After blinding, the user sends the deposit voucher [c', $v_i$, tx_id, $sign_f(c' \parallel v_i \parallel tx\_id)$] to the bank. This signature verifies the DT with tx_id to be sent by the user.

**Step 4. Bank performs the blind signature.**

The bank receives the depositing voucher and makes the following verification: ① whether or not the DT with tx_id has been used previously; ② whether or not the DT has acquired sufficient acknowledgements through the Bitcoin network. If these two verification steps are not passed, then the bank returns an error.

The bank collects the user's public key and the amount from the DT, and then checks the correctness of the signature and whether or not the amount is sufficient. If the signature is not correct, the bank returns an error. If the amount is not sufficient, the bank sends the Bitcoins back to the user's input address. The bank calculates s' = k − c'd, sends s' to the user, and saves c', $v_i$, and tx_id.

**Step 5. User unblinds the blind signature.**

The user receives s' and calculates s = s' + γ. The signature on m is (c, s) and (m, c, s) is the withdrawal voucher or payment voucher. The user verifies the signature by means of (F2). If (F2) is false, then an error is output.

**Step 6. Verification and withdrawal**

When the user withdraws from the bank or pays to a business, (m, c, s) is provided. The bank verifies: ① whether or not the message is correct, such as having the correct O and $v_i$, and whether or not the public key P equals that of the bank; ② whether or not (F2) is true; ③ whether or not the (m, c, s) has been used previously. If these three verification steps are not passed, then the bank returns an error.

If (m, c, s) is correct and the user wants to withdraw Bitcoins, the bank checks whether or not the withdrawal delay has been reached. If it has not been reached, an error is returned; otherwise, the bank randomly selects one of the withdrawal addresses containing Bitcoins not less than $v_i$, the bank sends $v_i$ Bitcoins to the user's output address O, which represents a transaction denoted as a withdrawal transaction (WT). If all existing withdrawal addresses contain insufficient Bitcoins, new withdrawal addresses should be generated, and the bank should send sufficient Bitcoins to these addresses from previous deposit addresses, which are denoted as dispatch transactions (PTs).

*3.4. Blind-Mixing Scheme Analysis*

*3.4.1 Untraceable*: the bank ensures that the user's input address and output address are untraceable through separating the deposit address from the withdrawal address and the withdrawal delay; thus, these two addresses are not connected with each other in the transaction chain. An attacker knows the input address of User 1, and traces transactions through the blockchain of the Bitcoin system from the DT to the PT and to the WT of User 2, as illustrated in Fig. 4. As such, the attacker could obtain the output address of User 2 through the input address of User 1; however, the output address of User 1 cannot be traced in this manner. Furthermore, because of the withdrawal delay, the attacker cannot deduce whether or not the WT of User 1 has occurred, or when it is scheduled to occur.

*3.4.2 Address management*: the deposit address and the withdrawal address of the bank appear twice in the blockchain, once when receiving Bitcoins and once when sending Bitcoins; however, the two addresses of a particular user are not related to each other. Thus, attackers cannot determine whether or not this is a coin-mixing transaction of the bank. Moreover, while a user participating in the coin-mixing transaction knows their own deposit and withdrawal addresses, and knows those of the other users, the deposit and withdrawal addresses cannot be linked together. In addition, using CoinJoin in dispatch transactions makes the dispatching process from the deposit address to the withdrawal address more anonymous.

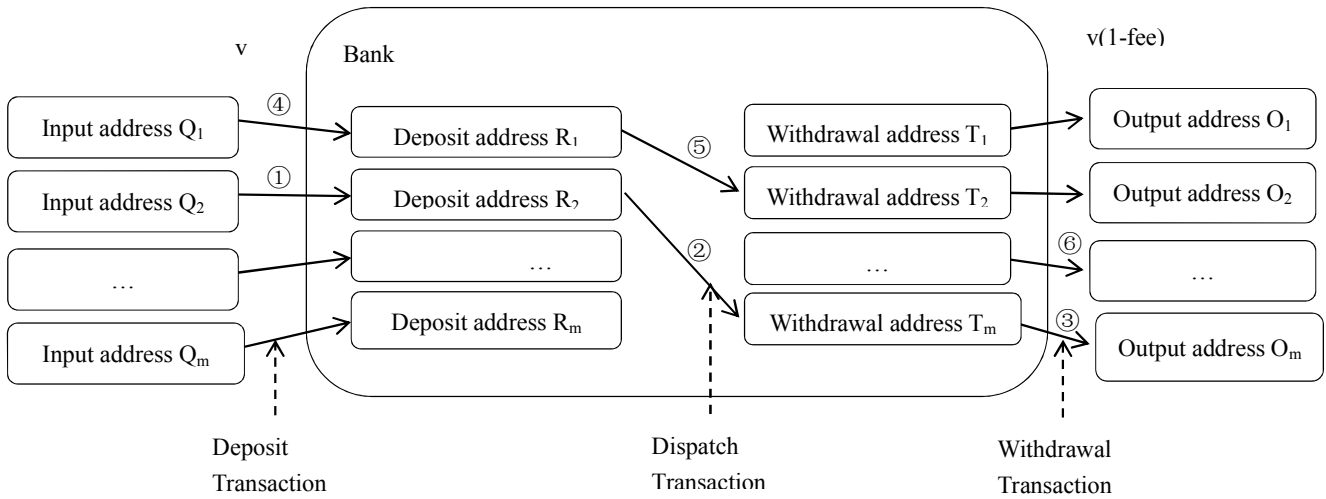

*Fig. 4* The three types of transactions and four types of addresses in Blind-Mixing

*3.4.3 Transaction denominations*: an attacker cannot determine the input and output addresses of users by tracing the input amount according to the denomination employed. For instance, if user $U_1$ sends 0.2 BTC to the bank while other users send 0.1 BTC, then it is easy to identity $U_1$'s input and output addresses. However, if $U_1$ also sends 0.1 BTC, then the opposite is the case. Transaction denominations affect the anonymity of coin-mixing. Suppose that the number of coin-mixing events is M and the number of denominations is N, then the average number of coin-mixing events of a certain denomination is M/N, and the probability of finding a particular user's output address is N/M. For only a single denomination, the probability of deanonymizing is 1/M, and, if M = N, then the probability will be 1, that is, the user's output address must be discovered. It is therefore best when N = 1.

*3.4.4 Unlinkablity*: the bank receives the deposit voucher before it is signed and obtains the withdrawal voucher after verification and withdrawal. There is only one commonality between the deposit voucher and the withdrawal voucher, which is the amount. However, the bank cannot successfully deanonymize based solely on the amount, and it must deduce the blind signature from the original message and the signature to link the signature and the blind signature. Nevertheless, to do so, the bank must know the two integers γ and δ, when, in fact, the bank knows nothing about them.

*3.4.5 Runtime or time-delay*: the DT and the WT have two methods of implementation. The first method is runtime transaction, where the bank sends Bitcoins to the user's output address immediately when the user requests withdrawal. The advantages of the method include ① friendly to users, and the WT requires very little time; ② does not uncover the output addresses of other users, and the WT is obfuscated by normal transactions within the blockchain. The weaknesses of the method include a greater number of transactions and more transaction fees. The second method is CoinJoin transactions with time-delay, where the bank places all user withdrawal requests together into a single CoinJoin transaction with numerous inputs and outputs. The advantages of the method include fewer transactions and transaction fees. The weaknesses of the method include ① the WT requires a certain time-delay; ② transactions are implemented with CoinJoin, where other users' input and output addresses are observed. Hence, the first method should be adopted for a relatively small number of coin-mixing activities, such that the coin-mixing transactions of other users can be hidden. As the number of coin-mixing activities increases, CoinJoin transactions with time-delay become an increasingly better method.

*3.4.6 Undeniability*: the bank may deny the public key of the deposit address provided by it so that it can deny that it ever received the user's Bitcoins. Therefore, the bank signs the public key R in Step 2, which proves that the public key is that provided by the bank. The blind signatures are also undeniable for the bank. Suppose there is a second private key $d_2$ that could generate the valid blind signature $(c', s_2')$, which provides the signature $(c, s_2)$ after unblinding. According to (F2), we obtain the following equation.

$c = SHA256(m \parallel R_x(cP + sG) \bmod n) = SHA256(m \parallel R_x(cP + s_2 G) \bmod n)$

…

$c = SHA256(m \parallel x \bmod n) = SHA256(m \parallel x_2 \bmod n)$

$c = SHA256(m \parallel t) = SHA256(m \parallel t_2)$

If $t \neq t_2$, there are two points of A, which contradicts known conditions. Therefore, $t \equiv t_2$, $(c, s) = (c, s_2)$, and $d = d_2$.

*3.4.7 Deposit voucher*: ECC with elliptic curve secp256k1 is adopted in Blind-Mixing, so the key pair in the Bitcoin system is compatible with that of Blind-Mixing. Therefore, while the user requests the bank to perform blind signature, the user must send the deposit voucher and the signature on it to the bank, where the private key f is from the input address in the transaction tx_id. The bank also obtains the public key Q from this input address, and then verifies the signature and the deposit voucher using the ECC signature algorithm. If the signature is correct, then the Bitcoins lodged at the input address belong to the user.

*3.4.8 Analysis of attacks*: anonymity of a subject means that the subject is not identifiable within a set of subjects, denoted as the anonymity set [22]. As for Blind-Mixing, the subject refers to a coin-mixing action of users. A passive attacker deanonymizes through analysis of the transaction chain. Suppose that an ordinary user $U_2$ joins Blind-Mixing, the mixing fee rate is *fee*, and the passive attacker knows the user's input address and wants to deduce the output address. As shown in Fig. 4, the passive attacker obtains information regarding three transactions from the Bitcoin blockchain: ① the DT of $U_2$; ② the PT from the deposit address of $U_2$ to the withdrawal address of user $U_m$; and ③ the WT of $U_m$. As such, the attacker deduces the deposit address of $U_2$, and the withdrawal address and the output address of $U_m$; however, the passive attacker cannot locate the withdrawal address of $U_2$.

The passive attacker also narrows the anonymity set through information regarding the Bitcoin amount and the timing sequence. Suppose the WT is located in a particular block within the $\omega$ blocks after the DT, and, within this range of transactions, M transactions with an input amount between $v(1-\text{fee})$ and $v$ may be the withdrawal transaction. The number M includes the number of normal transactions and the number of Blind-Mixing transactions.

In addition to the activities of a passive attacker, an active attacker takes part in Blind-Mixing activities. Given that $U_2$ is the active attacker and obviously knows his own input and output addresses, $U_2$ also obtains information regarding another three transactions in addition to the three transactions known by the passive attacker: ④ the DT of user $U_1$; ⑤ the PT from the deposit address of $U_1$ to the withdrawal address of $U_2$; ⑥ the WT of $U_2$. Therefore, as shown in Fig. 4, $U_2$ can deduce the deposit address of $U_1$ and the output address of $U_m$; however, $U_2$ cannot locate the withdrawal address of $U_1$ as well.

50 percent attack: the active attacker can obtain the deposit address of $U_1$ and the output address of $U_m$ through a single Blind-Mixing transaction. If the active attacker engages in half-Blind-Mixing activities, then the attacker can obtain all the input addresses $Q_i$ and output addresses $O_i$ of all the users, and all the deposit addresses $R_i$ and withdrawal addresses $T_i$ of the bank, and all the transactions (symbolically indicated by $\rightarrow$) related to Blind-Mixing. This is described in relation to the maximum information set $U_a$ that the active attacker obtains as follows:

$$U_a = \{Q_i \rightarrow R_i, R_i \rightarrow T_j, T_i \rightarrow O_i, 0 \leq i, j \leq M\}. \tag{F4}$$

However, even if the attacker has the ability to cheat the bank, the active attacker still cannot link $R_i$ with $T_j$.

The super attacker has some special abilities, which can be described as follows.

(1) Sniffing communication data. Blind-Mixing runs over TOR, so the super attacker cannot decode the TOR data.
(2) Control of the bank. The attacker can obtain no more information than that given by (F4). Because of the blind signature, the super attacker cannot link the input and output address of all users.

(3) Link DTs to WTs through the two IP addresses employed by the user when depositing and withdrawing. However, Blind-Mixing runs over TOR, and the IP address is always different. As such, deanonymization based on IP addresses is very difficult.

Clearly, Blind-Mixing resists the above attacker models.

## 4. Experiments and Analysis

We performed two experiments. The first experiment compares the performance of RSA and ECC, and the second involves a coin-mixing experiment of the Blind-Mixing system. We also compared Blind-Mixing with two other coin-mixing schemes based on blind signature, BlindCoin, and RSA Coin-Mixing.

*4.1. Comparison of Performance*

We implemented the ECC blind signature algorithm [18] with a 256-bit key size and the RSA blind signature algorithm with a 1024-bit key size [16] under Windows 8 running on a notebook with Core i7-5500U @ 2.40 GHz CPU and 16 G memory, and divided the overall process into 5 phases, which are initiation, blinding, blind signing, unblinding, and verification, to compare the performances of two algorithms in each phase. Each algorithm was run 10,000 times while collecting relevant data, which was repeated 10 times to obtain average values. The results are shown in Fig. 5.

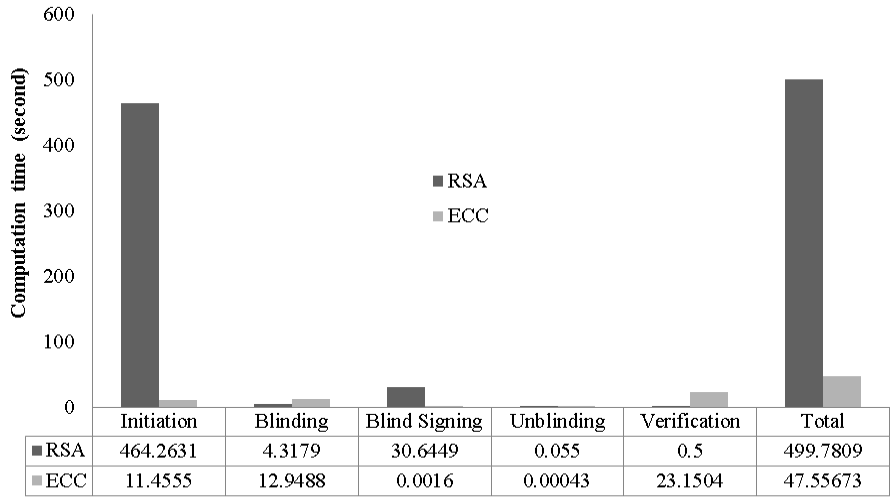

*Fig. 5* Comparison of the computation times of RSA and ECC blind signature algorithms during each of the 5 phases

The computation time of the initiation phase of RSA to generate 10,000 RSA public keys is quite large; however, the ECC algorithm employs a random large integer as the private key, so the speed of the initiation phase is 40 times greater than that of RSA. The performances of ECC for blind signing and unblinding are more efficient than those of RSA; however, the performances of blinding and verification require more computation time than those of RSA. In total, the ECC algorithm executes 10.5 times faster than RSA. The entire of RSA blind algorithm process required 0.05 s to execute, whereas the entire ECC process required only 0.00476 s.

*4.2. Coin-mixing experiment of the Blind-Mixing system*

We implemented a Blind-Mixing system based on the Blind-Mixing scheme. The experiment included two roles, which are the client and the bank. The client actions are requesting a public key, depositing, blind signing, verification, and withdrawal, and the bank returns the public key of the deposit address, blind signature, verification result, and withdrawal result, and also sends withdrawal transactions.

We performed Blind-Mixing on the Blind-Mixing system 10 times using a mixing amount of 0.001 BTC, and the bank charges 0.0001 BTC as a mixing fee. The client and the bank were run under Windows 7 on the same local network. The procedure is as follows.

(1) The client generates 10 input addresses and 10 output addresses, and 0.0011 BTC is sent to each input address (see the transaction in the blockchain website: https://blockchain.info/tx/571063021a60c785452408238f36f2916b4477049209dc33238c80f539c451c8).

(2) The bank generates 10 deposit addresses and 10 withdrawal addresses, and 0.0011 BTC is sent to each withdrawal address (see transaction https://blockchain.info/tx/f5fd9b62212f28f25dae2f08bc92a9f30f48d3e399cc690a9a11dd4b6645ce7a).

(3) Depositing: 0.001 BTC is respectively deposited from 10 input addresses to 10 deposit addresses, where all transactions are in the $370,503^{rd}$ block.

(4) Unblinding: blind signing and withdrawal are conducted according to the Blind-Mixing scheme. 0.0009 BTC is respectively withdrawn from 10 withdrawal addresses to 10 output addresses, and the bank retains 0.0001 BTC as the mixing fee. The withdrawal transactions are in the $370,504^{th}$ block and the $370,505^{th}$ block.

(5) Dispatching: 0.01 BTC is dispatched from 10 deposit addresses to 10 new withdrawal addresses using the CoinJoin method (see transaction https://blockchain.info/tx/5b8f2943e0ef7057b945782e3237c5a55513bab948de2ea83fc33594aa618173).

The transactions of the experiment on the Blind-Mixing system are shown on Fig. 6, and it is apparent that the transactions of the Blind-Mixing algorithm have fully completed, which verifies the feasibility of Blind-Mixing.

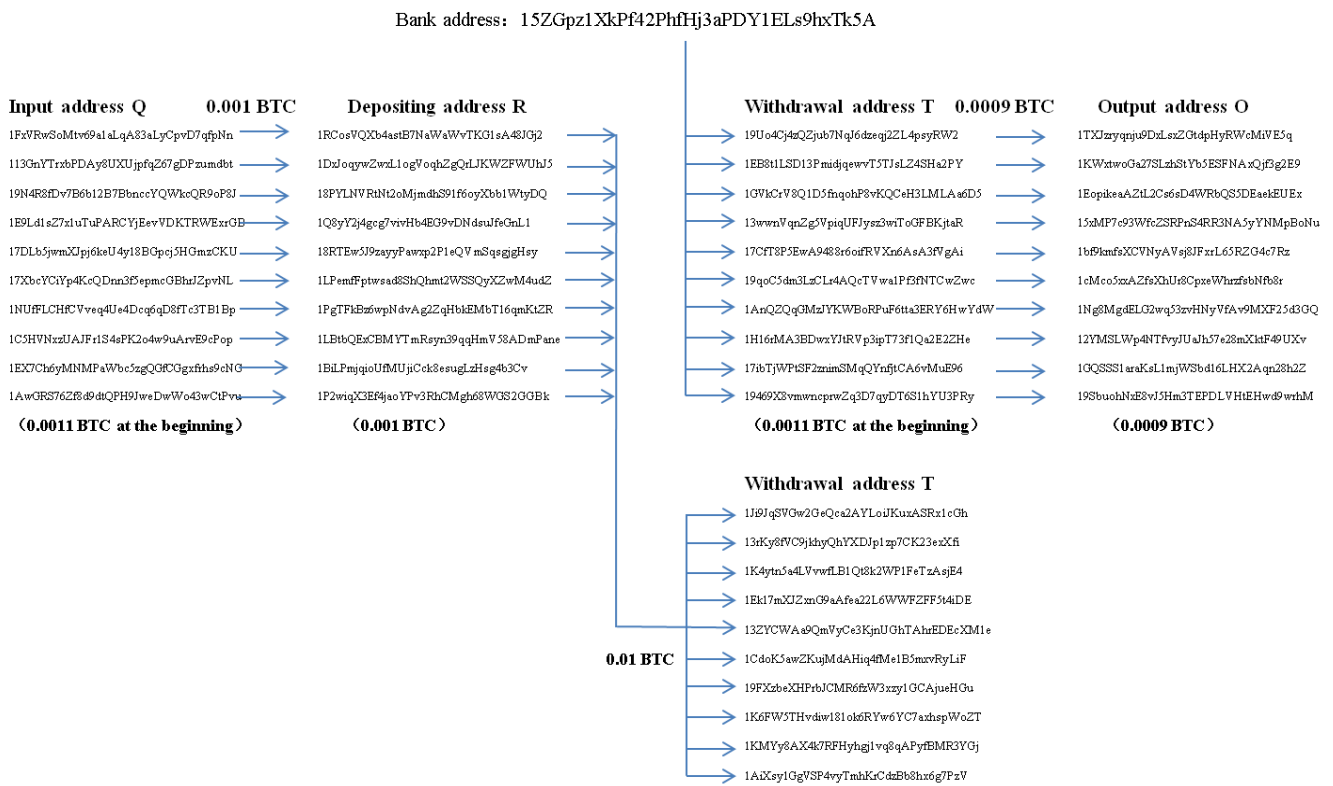

***Fig. 6*** Experimental transactions of the Blind-Mixing system

*4.3. Analysis and Comparisons*

Comparisons between Blind-Mixing, BlindCoin and RSA Coin-Mixing algorithms are listed in Table 1.

According to the public key cryptography algorithm, Blind-Mixing adopts ECC with a short key size and high performance, BlindCoin adopts double linear groups, and RSA Coin-Mixing adopts RSA with a long key size and low performance.

A user's deposit address and withdrawal address are unique for Blind-Mixing; thus, the deposit address and withdrawal address of another user are not discoverable. The deposit address of BlindCoin is also unique; however, RSA Coin-Mixing adopts a single public address of the bank, so attackers can easily obtain the input addresses and output addresses of all users.

The deposit voucher in Blind-Mixing requires a signature and a unique deposit address, and BlindCoin requires the deposit address only; however, in RSA Coin-Mixing, a user's Bitcoins may be falsely claimed by another person. Because transaction IDs are easily obtained by checking the public address of the bank, an active attacker can send a request for blind signature before the user does.

Table 1 Comparisons between Blind-Mixing, BlindCoin, and RSA Coin-Mixing algorithms

| Schemes | Blind-Mixing | BlindCoin | RSA Coin-Mixing |
|---|---|---|---|
| Cryptography | ECC | Double linear groups | RSA |
| Performance of Cryptography | Good | Unknown | Normal |
| The deposit address | Unique | Unique | Public address |
| The withdrawal address | Unique | Unknown | Public address |
| Deposit verification | Deposit address and signature | Deposit address | Transaction ID |
| Weakness | None | Public log | Public address and deposit verification |
| Attacker model | Resist super attacker | Resists passive attacker | Resists passive attacker |
| Feasibility | Implemented | In theory | Implemented |

BlindCoin publishes user information in a public log, and thus supports third-party verification and accountability; however, this weakens the anonymity of BlindCoin, and an active attacker may conduct the following actions.

(1) Join the normal process of coin-mixing, obtain the block difference α between a DT and the time when the blind signature information is published, and obtain the block difference β between a WT and the time when the signature information is published.

(2) When the bank receives the Bitcoins from a user deposit, it publishes the blind signature information on the public log, and exposes the DT, which typically occurs α blocks prior to the publishing time with an input amount of v.

(3) When the user makes a withdrawal and publishes signature information on the public log, the DT typically occurs β blocks after the publishing time with an input amount of v.

(4) By combining knowledge of α, β, and v, there is a high possibility of deducing the DT and the WT, and then obtaining the input and output addresses.

Therefore, BlindCoin and RSA Coin-Mixing cannot resist active attackers; however, Blind-Mixing can resist even super attackers. In addition, the Blind-Mixing system has already been implemented and has performed well in experiments, but the BlindCoin scheme remains theoretical.

## 5. Conclusions

This paper proposed a centralized coin-mixing algorithm denoted as Blind-Mixing based on the elliptic curve blind signature scheme, which avoids conditions where mixers can obtain the input and output addresses of users. Thus, the proposed scheme improves the anonymity of centralized coin-mixing. The paper also considered three types of attacker models and their effects on Blind-Mixing.

The results of the analysis showed that Blind-Mixing can resist even super attackers. In addition, BlindCoin may be deanonymized successfully with high possibility because of its use of a public log. Moreover, in RSA Coin-Mixing, a user's Bitcoins may be falsely claimed by another, and other users' deposit and withdrawal transactions are easily exposed. Therefore, the comparisons conducted show that Blind-Mixing provides for better anonymity than BlindCoin and RSA Coin-Mixing.

This paper also compared the performances of the RSA and ECC blind signature algorithms, and the results showed that the ECC blind signature scheme provides an overall computational speed that is 10.5 times greater than that of the of RSA blind signature scheme. We implemented a Blind-Mixing system based on the Blind-Mixing scheme, and testing verified that the Blind-Mixing system is feasible for practical applications.